\documentclass[aps,superscriptaddress,footinbib,twocolumn]{revtex4} %preprint
\usepackage{dsfont}
\usepackage{amsmath}
\usepackage{graphicx}
\usepackage[latin1]{inputenc}
\usepackage{version} % for comment
\usepackage{color}
\usepackage{amsmath}
\usepackage{colortbl}
\usepackage{xcolor}  % for writting parts in color, e.g. in red
\usepackage[colorlinks=true,citecolor=black,urlcolor=black,linkcolor=black]{hyperref} % choice of color for hyperref links
\makeindex
\begin{document}

\title{Local emergence of thermal correlations in an isolated quantum many-body system} 

\author{T. Langen}
\email[]{tlangen@ati.ac.at}    
\affiliation{Vienna Center for Quantum Science and Technology, Atominstitut, TU Wien, Stadionallee 2, 1020 Vienna, Austria}
\author{R. Geiger}   
\affiliation{Vienna Center for Quantum Science and Technology, Atominstitut, TU Wien, Stadionallee 2, 1020 Vienna, Austria}
\author{M. Kuhnert}   
\affiliation{Vienna Center for Quantum Science and Technology, Atominstitut, TU Wien, Stadionallee 2, 1020 Vienna, Austria}
\author{B. Rauer}   
\affiliation{Vienna Center for Quantum Science and Technology, Atominstitut, TU Wien, Stadionallee 2, 1020 Vienna, Austria}
%\author{\red{M. Gring(?)}}   
%\affiliation{Vienna Center for Quantum Science and Technology, Atominstitut, TU Wien, Stadionallee 2, 1020 Vienna, Austria}
\author{J. Schmiedmayer}
\email[]{schmiedmayer@atomchip.org}
\affiliation{Vienna Center for Quantum Science and Technology, Atominstitut, TU Wien, Stadionallee 2, 1020 Vienna, Austria}

%------------------------------- abstract
\label{par:abstract}
\begin{abstract}
We experimentally demonstrate how thermal properties in an non-equilibrium quantum many-body system emerge locally, spread in space and time, and finally lead to the globally relaxed state. In our experiment, we quench a one-dimensional (1D) Bose gas by coherently splitting it into two parts. By monitoring the phase coherence between the two parts we observe that the thermal correlations of a prethermalized state emerge locally in their final form and propagate through the system in a light-cone-like evolution. Our results underline the close link between the propagation of correlations and relaxation processes in quantum many-body systems.
\end{abstract}
%------------------------------- end abstract

%\date{\vspace{-5ex}}
\date{\today}

\maketitle

\label{par:intro}

%------------------------------- intro
Understanding the dynamics of isolated quantum many-body systems is a central open problem at the intersection between statistical physics and quantum physics~\cite{polkovnikov}. It has been theoretically suggested that relaxation in generic isolated quantum many-body systems proceeds through the dephasing of the quantum states populated at the onset of the non-equilibrium evolution~\cite{rigol,srednicki}. It is generally believed that this dynamically leads to relaxed states which can be well described either by the usual thermodynamical ensembles, or by generalized Gibbs ensembles which take into account dynamical constraints~\cite{Rigol2007}. However, it remains an open question how these relaxed states form dynamically, and in particular, whether they emerge gradually on a global scale, or appear locally and then spread in space and time. 

Ultracold atomic gases offer an ideal test bed to explore such quantum dynamics. Their almost perfect isolation from the environment and the many available methods to probe their quantum states make it possible to reveal the dynamical evolution of a many-body system at a very detailed level~\cite{kinoshita,gring,gaunt,cheneau,sadler,ritter,trotzky,gerving}.

\begin{figure}[h!tb]
	\centering
		\includegraphics[width=0.46\textwidth]{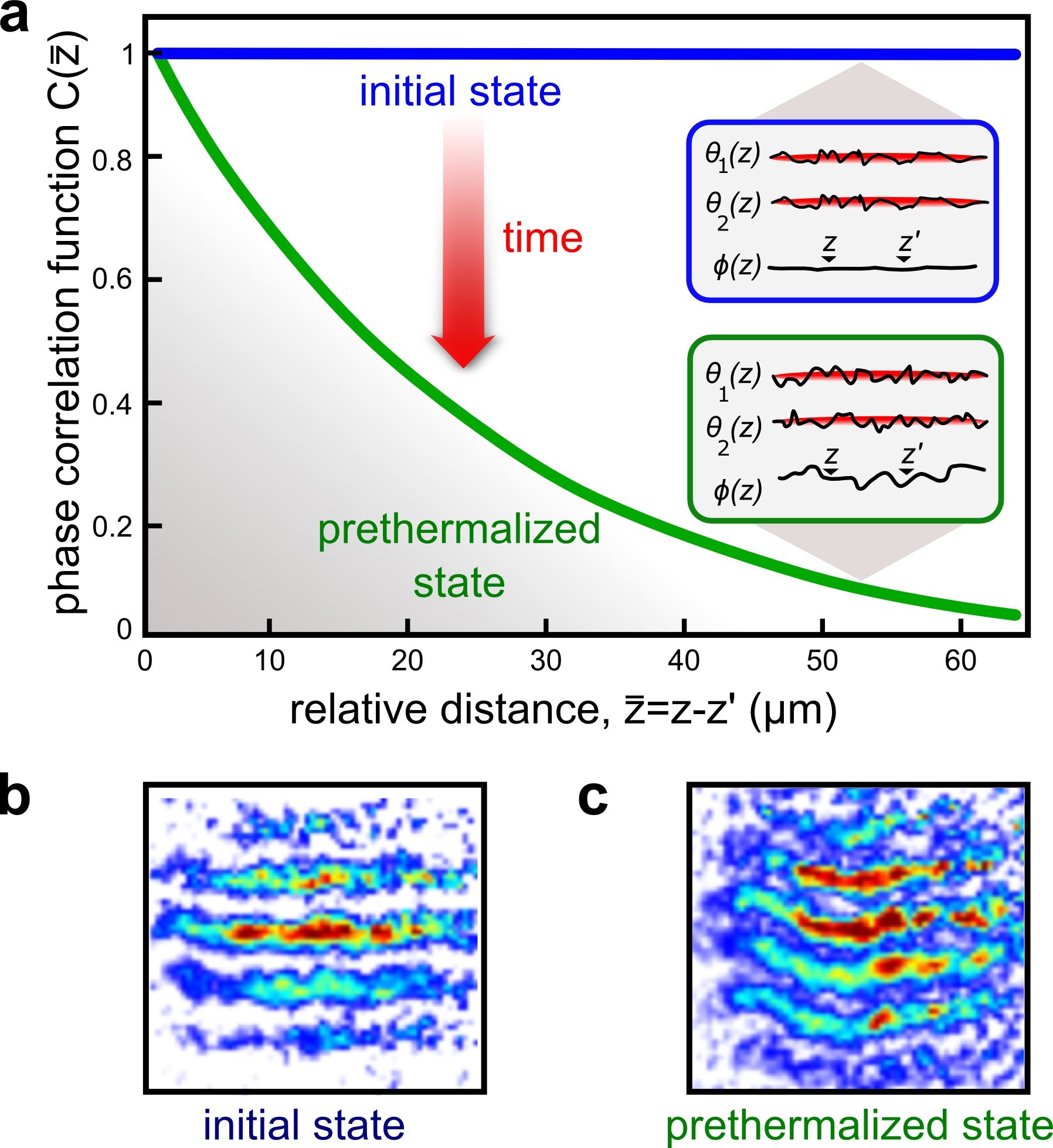} %475
	\caption{\textbf{Characterizing the dynamics of correlations in a coherently split 1D Bose gas.}
	\textbf{(a)} The splitting process creates two 1D gases with almost identical longitudinal phase profiles $\theta_1(z)$ and $\theta_2(z)$, corresponding to long-range phase coherence in the relative phase field $\phi(z)=\theta_1(z)-\theta_2(z)$. 
	The degree of relative phase correlations between two arbitrary points $z$ and $z'$ along the length of the system is characterized by the two-point correlation function $C(\bar z,t)$. Initially, it is close to unity for any distance $\bar z = z-z'$ between the points.
	Over time, this strongly phase-correlated state relaxes towards a prethermalized state, characterized by thermal (exponentially decaying) correlations.
	The aim of this study is to investigate how the thermal correlations locally emerge in time. 
In the experiment, the relative phase field is probed via matter-wave interferometry between the two gases. \textbf{(b,c)} show example interference pictures in the initial and in the prethermalized state, respectively. In these pictures, the relative phase $\phi(z)$ is directly extracted from the local position of the interference fringes. The phase correlation function is then calculated from an average over approximately $150$ interference pictures.%\vspace{-35pt}
}
	\label{fig:figure1}
\end{figure}

\label{par:experiment}
In our experiment, a phase-fluctuating ultracold 1D Bose gas~\cite{petrov} is split coherently~\cite{schumm}. The splitting creates a non-equilibrium state consisting of two gases with almost identical phase profiles. Interactions in the many-body system drive the relaxation of this highly phase-correlated state to a prethermalized state, characterized by thermal phase correlations~\cite{gring,kuhnert}. The dynamics is monitored by time-resolved measurements of the relative phase field using matter-wave interferometry~\cite{cronin}. 

%------------------------------- end intro

%------------------------------- experiment
The experimental procedure starts with a 1D degenerate gas of $4000-12000$ ${}^{87}$Rb atoms trapped at temperatures between $30-110\,$nK in a magnetic trap, formed $100\,\mu$m below the trapping wires of an atom chip~\cite{reichel}. By applying radio-frequency fields via additional wires on the chip, we rapidly transform the initial harmonic trapping potential into a double well, thereby realizing the matter-wave analogue of a coherent beam splitter~\cite{schumm,Splitting}. The system is let to evolve in the double well for a variable time $t$, before the gases are released by switching off the trapping potential. They expand and interfere after a time-of-flight of $15.7\,$ms. The resulting interference pattern allows to extract the relative phase $\phi(z,t)=\theta_1(z,t)-\theta_2(z,t)$ along the length of the system (see Fig.~\ref{fig:figure1} and \cite{PhaseExtraction}). Here $\theta_1(z,t)$ and $\theta_2(z,t)$ are the phase profiles of the two individual gases. Repeating this procedure approximately $150$ times for each value of $t$, we determine the two-point phase correlation function 
\begin{equation}
{C}(\bar z = z-z^\prime,t) = \mathrm{Re }\,\langle e^{i \phi (z,t)-i\phi (z^\prime,t) }\rangle.
\end{equation}
It measures the degree of correlation between the phases at two arbitrary points $z$ and $z^\prime$, separated by a distance $\bar z$~\cite{betz,whitlock}. In contrast to the integrated visibility of the interference pattern, which was used in a previous experiment to identify the prethermalized state~\cite{gring,berges}, the phase correlation function provides a local probe for the dynamics, and is therefore ideally suited to study the propagation of correlations. 

%------------------------------- experimental results
\begin{figure}[tb]
	\centering
		\includegraphics[width=0.46\textwidth]{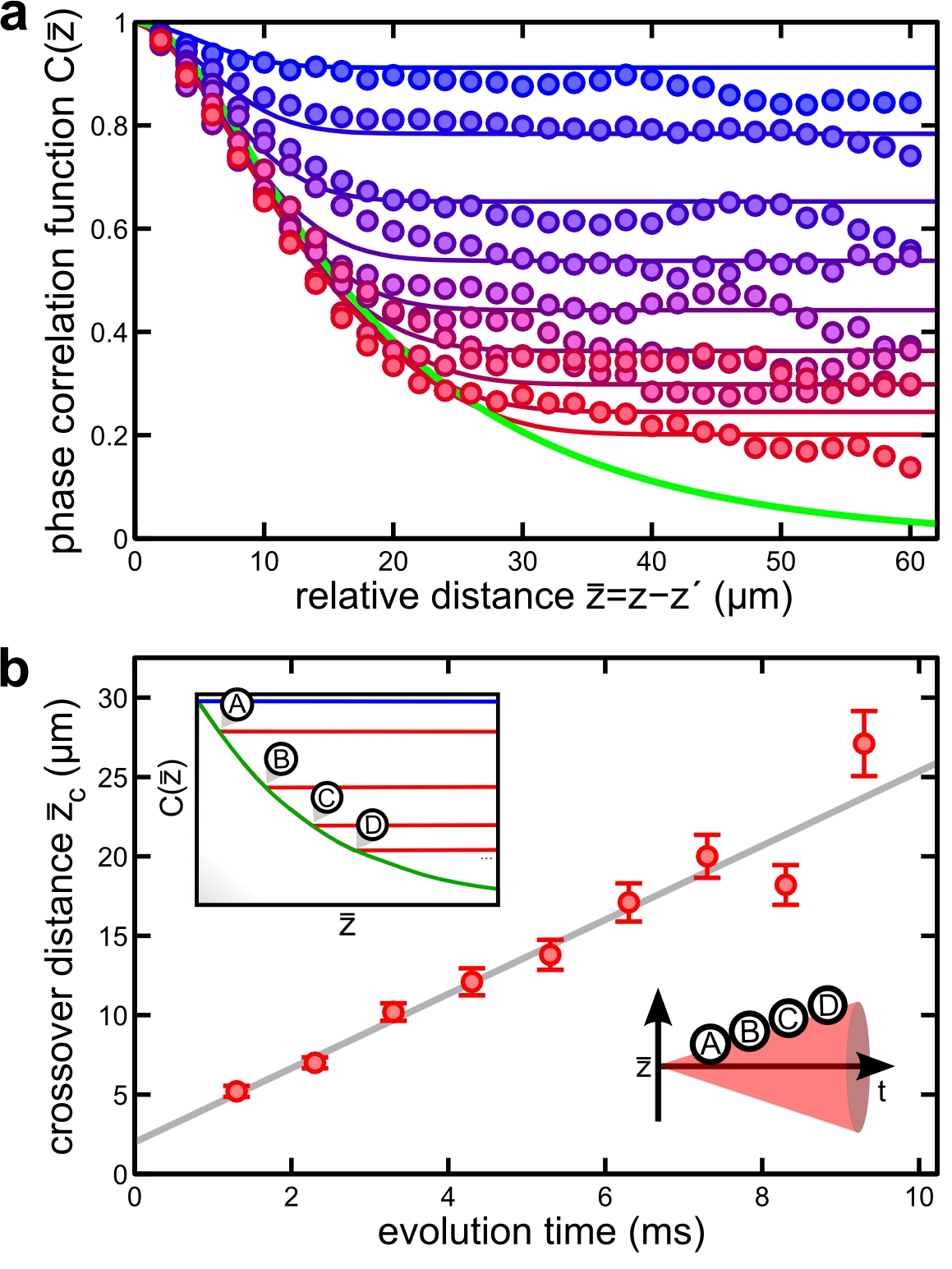}
	\caption{\textbf{Local emergence of thermal correlations in a light-cone-like evolution.} \textbf{(a)} Experimental phase correlation functions $C(\bar z,t)$ (filled circles) compared to theoretical calculations (solid  lines). From top to bottom, the evolution time $t$ increases from $1\,$ms to $9\,$ms in steps of $1\,$ms. The bottom (green) line is the theoretical correlation function of the prethermalized state. For each $t$, the constant values of $C(\bar z,t)$ at large $\bar z$ can be used to determine the crossover distance $\bar z_c(t)$ up to which the system forgets the initial long-range phase coherence (see text for details). \textbf{(b)} Position of the crossover distance $\bar z_c$ as a function of evolution time $t$, revealing the light-cone-like decay of correlations. The solid line is a linear fit, the slope of which corresponds to twice the characteristic velocity of correlations. Insets: schematic visualization of the dynamics. The decay of correlations is characterized by a front moving with a finite velocity: for a given time $t$, $C(\bar z,t)$ is exponential (thermal) only up to the characteristic distance $\bar z_c(t)$ (points A-D). Beyond this horizon, long-range phase coherence is retained. Note that in the experimental data shown in (a), the sharp transitions (points A-D) are smeared out by the finite experimental imaging resolution.
		}
	\label{fig:figure2}
\end{figure}

\label{par:results}
Typical experimental data is presented in Fig.~\ref{fig:figure2}a. Directly after the quench, the phase correlation function ${C}(\bar z,t)$ is close to unity for any distance $\bar z$. This is a direct manifestation of the long-range phase coherence produced by the splitting process. %, which prepares the two gases with nearly identical phase profiles (ref gring?). %As the two phase profiles are nearly identical, the relative phase between the two gases $\phi(z)$ is constant and exhibits only small fluctuations caused by the quantum noise introduced by the splitting process. 
After a given evolution time $t$, the phase correlation function decays exponentially up to a characteristic distance $\bar z_c$ and stays nearly constant afterwards: $C(\bar z>\bar z_c,t)=C(\bar z_c,t)$. This means that beyond the distance $\bar z_c$ long-range phase coherence is retained across the system. %, as revealed by the constant value of $C(\bar z,t)$ for $\bar z>\bar z_c$. 
With longer evolution time, the position of $\bar z_c$ shifts to larger distances and the value of ${C}(\bar z> \bar z_c,t)$ gradually decreases. The evolution continues until the system reaches a quasi-steady state, where the correlations decay exponentially throughout the entire system. This prethermalized state corresponds to the relaxed state of the 1D system and can be described by a generalized Gibbs ensemble~\cite{gring,Rigol2007}.

From the experimental data, we extract the crossover points $\bar z_c$ through the level of long-range phase coherence. To this end, we consider for each $t$ the region where the correlation function is constant, extrapolate the constant value to smaller $\bar z$ and determine the position $\bar z_c$ where it crosses the prethermalized correlation function. The result of this procedure is shown in Fig. 2b. We observe a clear linear scaling of the position $\bar z_c = 2ct$, characterizing the local decay of correlations with time. This observation reveals that an arbitrary point in the gas loses its correlations with other points up to a certain separation $\bar z_c$, while long-range phase coherence persists outside this horizon. The experimental data thus show that the prethermalized state locally emerges in a light-cone-like evolution, where $c$ plays the role of a characteristic velocity for the propagation of correlations in the quantum many-body system. For the data presented in Fig.~\ref{fig:figure2}b a linear fit allows to extract a velocity of $c=1.2 \pm 0.1\,$mm/s.

%------------------------------- theory
\label{par:explanation}
Light-cone-like effects in quantum many-body dynamics have been previously predicted using results from conformal field theory~\cite{calabrese}, and for 2D superfluids~\cite{mathey}. Similarly, it is known that some microscopic lattice models exhibit an intrinsic maximum velocity~\cite{lieb}, which limits the propagation of correlations and entanglement to an effective light-cone \cite{bravyi,cramer,laeuchli}. However, a direct connection to the relaxation of continuous quantum many-body systems and the establishment of thermal properties has, so far, not been observed.
  
The light-cone like emergence of thermal correlations which we observe in this work, can be understood using a homogeneous Luttinger Liquid (LL) model that effectively describes the interacting many-body system in terms of low-energy excitations~\cite{giamarchi}. Within the LL model, these excitations are superpositions of phase and density fluctuations. They are characterized by a linear dispersion relation $\omega_k=c_0 |k|$, with $k$ being the momentum of the excitation and $c_0$ the speed of sound, the latter defining the characteristic velocity in the homogeneous system.

The coherent splitting process equally distributes energy among the excitations, resulting in a $1/k$ dependence of their occupation numbers~\cite{kitagawa}. Each excitation is initialized with small relative phase fluctuations and high relative density fluctuations. Over time, the amplitude of the phase (density) fluctuations increases (decreases), resulting in a progressive randomization of the relative phase field $\phi(z)$. Eventually, the energy associated with the phase fluctuations equilibrates with the energy associated with the density fluctuations, leading to the thermal phase correlations of the prethermalized state~\cite{kitagawa}. 

For a given evolution time $t$, the dephasing of the excitations with different wavelengths ($2\pi/k$) randomizes the relative phase field only up to a characteristic distance $\bar z_c=2c_0t$.  
This effect can intuitively be understood in the following way: the degree of randomization of the phase is
related to the amplitude of the contributing phase fluctuations. For large distances they are associated
with the highly occupied long wavelength excitations which take a long time ($\sim 1/\omega_k$) to be converted 
from the initial density fluctuations into phase fluctuations. 
At time $t$, there exists a characteristic distance beyond which the contribution of these long-wavelength fluctuations to the randomization of the phase is compensated by a decrease of the contribution from the faster short-wavelength fluctuations.
Therefore, the phase does not randomize any further and long-range phase coherence remains beyond $\bar z_c$. The sharpness of the transition at $\bar z_c$ results from the interference of the many excitations with different momenta.

In a more mathematical formulation, the phase correlation function can be written as $C(z,z',t)= \exp(-\frac{1}{2}\langle\Delta\phi_{zz'}(t)^2\rangle)$. In the homogeneous limit, the local phase variance is given by~\cite{kitagawa,bistritzer,langen}
\begin{equation}
	%\langle\phi(z,z')^2\rangle \sim \sum_{k\neq0} \frac{\sin(\omega_k t)^2}{k^2}\left(1-\cos(kz)\right).
	\langle\Delta\phi_{zz'}(t)^2\rangle = \frac{2\pi^2}{L K^2} \sum_{k\neq0} \frac{\sin(\omega_k t)^2}{k^2}\left(1-\cos(k\bar z)\right),
	\label{eq:sum}
\end{equation}
with $L$ being the length of the system, $k=2\pi n/L$ the momentum of the excitations ($n\neq 0$ integer) and $K$ the Luttinger parameter. A very similar expression can be derived for the trapped system probed in the experiment. 

The first term in the sum \eqref{eq:sum} represents the growth and subsequent oscillations in the amplitude of the phase fluctuations as they get converted from the initial density fluctuations. The factor $1/k^2$ in the amplitude reflects the $1/k$ scaling of the excitation occupation numbers associated with the equipartition of energy induced by the fast splitting. The second term in the sum corresponds to the spatial fluctuations. Expression \eqref{eq:sum} is the Fourier decomposition of a trapezoid with a siding edge at $\bar z_c =2c_0t$, which explains the two step feature of the phase correlation function. 

\begin{figure}[tb]
	\centering
		\includegraphics[width=0.45\textwidth]{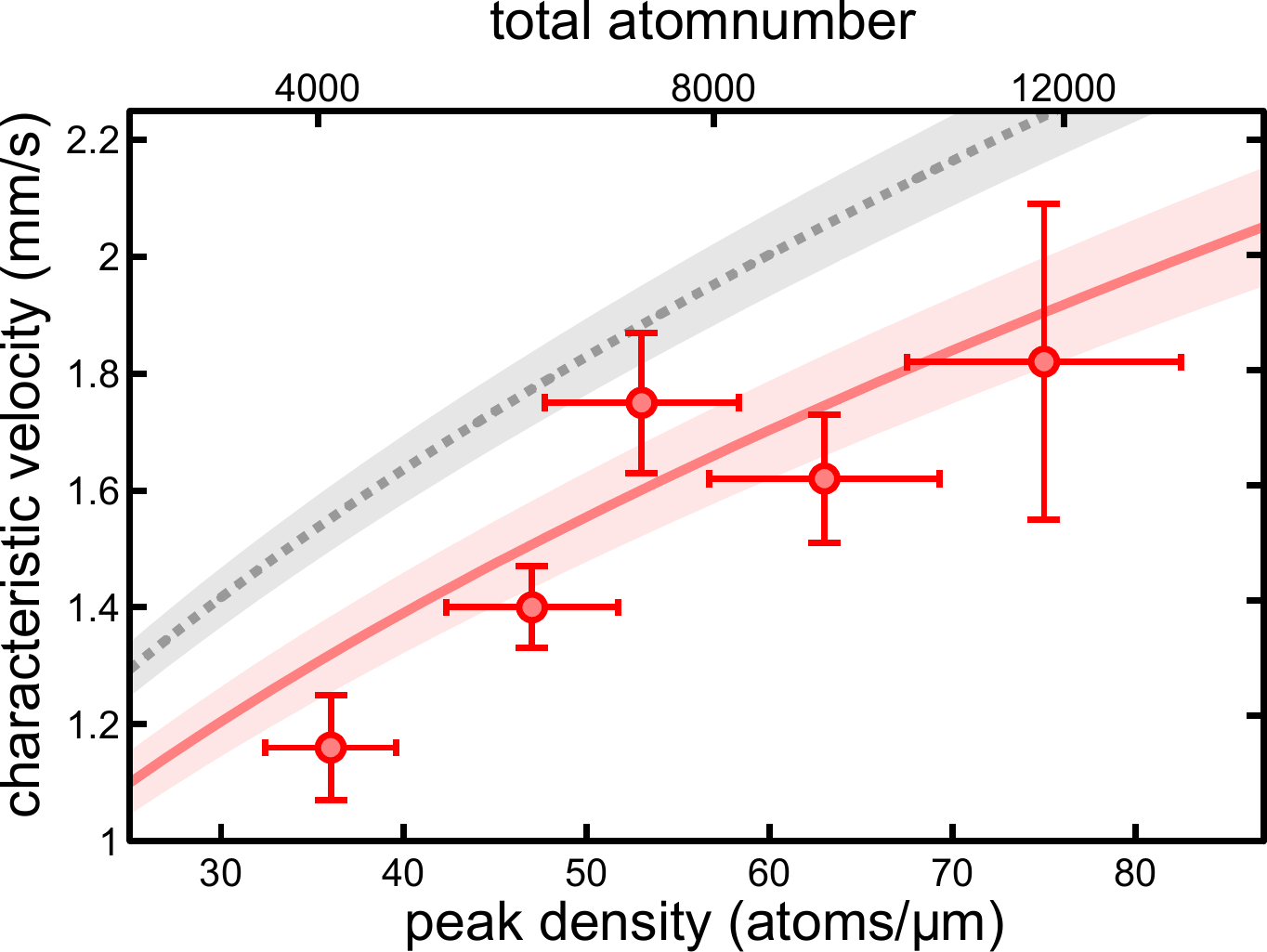}
	\caption{\textbf{Scaling of the characteristic velocity with particle number.} The solid red (dashed gray) line is the calculated velocity of correlations for a trapped (homogeneous) system. The peak densities  are given for each gas. Shaded areas correspond to the uncertainty on the measured trap frequencies. Error bars denote one standard deviation.
	}
	\label{fig:figure3}
\end{figure}

Alternatively, the excitations in the LL model can also be identified as pairs of quasi-particles, which propagate in opposite directions with momenta $k$ and $-k$, respectively. This interpretation naturally leads to the light-cone condition, as two points separated by $\bar z$ can establish thermal correlations if quasi-particles originating from these points meet after a time $t=\bar z/2c_0$. This interpretation thus shares strong similarities with theoretical results for the propagation of correlations and entanglement in systems close to a quantum critical point~\cite{calabrese} and with the recent experimental observation of a light-cone-like spread of correlations in a quenched lattice system \cite{cheneau}.

In Fig.~\ref{fig:figure2}a we compare the results of the LL calculation to our measured data, taking into account the finite resolution of the imaging system. We find good agreement, using independently measured experimental parameters as the input for the theory. This quantitative agreement validates our interpretation of the observations as the local emergence of thermal correlations.

To investigate the scaling properties of the characteristic velocity, we perform the experiment for a varying number of atoms $N$ in the system. We observe the light-cone-like emergence of the thermal correlations over the whole range of probed atom numbers ($N\sim 4000-12000$). In the experimentally realized trapped system, the speed of sound varies along the length of the system. Nevertheless, the superposition of many excitations still leads to a single characteristic velocity for the dynamics, which is slightly reduced with respect to the homogeneous case. In Fig. 3 we show the measured characteristic velocities. A LL calculation including the trapping potential describes the experimental data within the experimental error, whereas a purely homogeneous calculation clearly overestimates the characteristic velocity.

\label{par:conclusion}
 
In our experiment thermal correlations emerge locally. A local observer would see the final relaxed correlation function appear immediately after the splitting and spread through the system in a light-cone horizon-like fashion, while long-range phase coherence remains outside. This leads us to conjecture a general pathway to relaxation and the emergence of classical properties in isolated quantum many-body systems:  the decay of quantum coherence starts locally and then spreads through the system to establish a globally relaxed (dephased) state.  In systems where interactions manifest themselves in excitations with a linear dispersion relation the decay of quantum coherence takes the form of an effective light cone.

%\textbf{~\\Acknowledgements\\~\\}
\label{par:ack}
We would like to thank David Adu Smith and Michael Gring for contributions in the early stage of the experiment, Igor Mazets, Valentin Kasper and J\"urgen Berges for discussions and Jean-Fran\c cois Schaff and Thorsten Schumm for comments on the manuscript. This work was supported by the Austrian Science Fund (FWF) through the Wittgenstein Prize and the EU through the projects QIBEC and AQUTE.  T.L. and M.K. thank the FWF Doctoral Programme CoQuS (\textit{W1210}), RG is supported by the FWF through the Lise Meitner Programme M 1423.

%---------------------------------------

\label{biblio}
\bibliography{light_cone_biblio2}

\begin{thebibliography}{32}
\bibitem[{Spl()}]{Splitting}
\bibinfo{note}{The splitting is performed by linearly increasing the amplitude
  of the radio-frequency (RF) current in the chip wires to $24\,$mA within
  $12\,$ms. To minimize longitudinal excitations during the splitting, the
  initial gas is prepared in a slightly dressed RF trap that has the same
  longitudinal confinement as the final double well potential. The increase of
  RF current results in a rapid decay of the tunnel coupling between the two
  gases. Simulations of the chip potential and experiments with
  quasi-condensates in thermal equilibrium~\cite{betz} indicate that the
  decoupling of the two gases happens within less than $500\,\mu$s. This is
  faster than the characteristic timescale of the dynamics and therefore
  realizes a quench.}
\bibitem[{Pha()}]{PhaseExtraction}
\bibinfo{note}{The point spread function of the optical system has a measured
  rms width of $3.6\,\mu$m. The phase $\phi(z)$ of the interference patterns is
  extracted by fitting each pixel line (of size $\sigma_\mathrm{px}=2\,\mu$m)
  with a cosine-modulated Gaussian function.}
\end{thebibliography}

\end{document}